\documentclass[useAMS,usenatbib]{mn2e}

\usepackage{amsmath}									
\usepackage{graphicx} 									
\usepackage[english]{babel}								
\usepackage{subfigure}									
\usepackage{paralist}									
\usepackage{url}
\usepackage{color}

\begin{document}
	\title[Spectroscopic detection of reflected light]{Spectroscopic direct detection of reflected light from extra-solar planets}

	\author[J. H. C. Martins et al]{J. H. C. Martins$^{1,2}$,
        P. Figueira$^{1}$,
		N. C. Santos$^{1,2}$ and
		C. Lovis$^{3}$\\
        $^{1}$Centro de Astrof\'isica, Universidade do Porto, Rua das Estrelas, 4150-762 Porto, Portugal\\
		$^{2}$Departamento de F\'isica e Astronomia, Faculdade de Ci\^encias, Universidade do Porto, Rua do Campo Alegre, 4169-007 Porto, Portugal\\
		$^{3}$Observatoire de Gen\`eve, Universit\'e de Gen\`eve, 51 ch. des Maillettes, 1290, Versoix, Switzerland\\}
	\date{In original form: 2013 July 23; Accepted: 2013 August 29}
	\journal{Mon. Not. R. Astron. Soc., 2013 \hspace{10cm} (MN \LaTeX\ style file v2.2)}
	
	\maketitle
	\begin{abstract}
		At optical wavelengths, an exoplanet's signature is essentially reflected light from the host star - several orders of magnitude fainter. Since it is superimposed on the star spectrum its detection has been a difficult observational challenge. However, the development of a new generation of instruments like ESPRESSO and next generation telescopes like the E-ELT put us in a privileged position to detect these planets' reflected light as we will have access to extremely high signal-to-noise ratio spectra. With this work, we propose an alternative approach for the direct detection of the reflected light of an exoplanet. We simulated observations with ESPRESSO@VLT and HIRES@E-ELT of several star+planet systems, encompassing 10h of the most {favourable} orbital phases. To the simulated spectra we applied the Cross Correlation Function to operate in a much higher signal-to-noise ratio domain than when compared with the spectra. The use of the Cross-Correlation Function permitted us to recover the simulated the planet signals at a level above $3\,\sigma_{noise}$ significance on several prototypical (e.g., Neptune type planet with a 2 days orbit with the VLT at $4.4\,\sigma_{noise}$ significance) and real planetary systems (e.g., 55 Cnc e with the E-ELT at $4.9\,\sigma_{noise}$ significance). Even by using a more pessimistic approach to the noise level estimation, where systematics in the spectra increase the noise 2-3 times, the detection of the reflected light from large close-orbit planets is possible. We have also {{{shown}}} that this kind of study is currently within reach of current instruments and telescopes (e.g., 51 Peg b with the VLT at $5.2\,\sigma_{noise}$ significance), although at the limit of their capabilities.
	\end{abstract}
	

	\section{Introduction}
		Since the detection of a giant planet orbiting the solar-type star 51 Peg \citep{1995Natur.378..355M}, the number of known extrasolar planets keeps increasing steadily. At the moment of the writing of this paper, over 900 {{extra-solar}} planets in more than 700 planetary systems have since been published \citep[][\url{http://exoplanet.eu/}]{2011A&A...532A..79S}.

		The focus of extrasolar planet researchers is now divided in two main lines:
		\begin{inparaenum}[i)]
			\item the detection of lower and lower mass planets, with the goal of finding an Earth sibling and
			\item the detailed {{characterisation}} of planets orbiting other
		\end{inparaenum}
		stars. Both lines of research have proven very successful. In what concerns detection methods, both radial velocity \citep[e.g. with the HARPS instrument,][]{2009A&A...507..487M} and transit surveys \citep[e.g. with CoRoT and Kepler satelites,][]{2012ApJ...745..120B,2009A&A...506..287L} have been finding an increasing number of low mass/radius planets orbiting solar-type stars, some of {{{which are in}}} the so called Habitable Zone \citep[e.g.][]{2011A&A...534A..58P,2012ApJ...745..120B}. On the characterisation side, the precision of the transit measurements, in combination with the finest interior models, has now allowed to determine the bulk composition of the planetary structure of several planets, some of which seem to be mostly rocky/iron in nature \citep[e.g.][]{2011ApJ...729...27B,2009A&A...506..287L}. For the most {{favourable}} cases, exquisite photometric measurements have further allowed to detect both the emitted (IR) and reflected (optical) light of exoplanets using occultations or transmissions spectroscopy \citep[e.g.][]{2009A&A...501L..23A,2009Sci...325..709B,2011MNRAS.417L..88K,2012ApJ...751L..28D}.

		However, the detection of detailed spectral features in the atmosphere of other planets has remained one of the most challenging goals. Interestingly, recent campaigns have shown that such measurements are possible with present day instrumentation, even from the ground \citep[e.g.][]{2010Natur.465.1049S,2012Natur.486..502B,2012ApJ...753L..25R}. The detection of spectral features (e.g. spectral lines) from the planet atmosphere has a huge scientific value, no matter if we are measuring the planetary intrinsic emitted light, the atmosphere transmission spectrum, or the stellar reflected light on the planet's atmosphere.

		In particular, the detection of the reflected light spectrum would allow to determine the orbital velocity of the planet and thus its mass, like in the case of double-line eclipsing stellar binaries \citep[e.g.][]{2012Natur.486..502B,2012ApJ...753L..25R}. {Excluding scaling factors, the reflected spectrum of a planet can be seen as the stellar spectrum multiplied by an albedo function. For the sake of simplicity, this function is often assumed as constant over a given wavelength range, but a more realistic approach is to consider a wavelength dependent albedo function. Such a function can be constructed to represent the atmospheric properties of the planet and its spectral features and chemical composition \citep[e.g.][]{2002MNRAS.330..187C}. Therefore, the detection of a planet's reflected light can give critical information about its atmosphere and composition, which should help improve and constrain the physics of current atmosphere models. The reflected spectrum from the planet should even allow to retrieve its rotation rate} \citep{2012ApJ...760L..13K}. Unfortunately, such measurements are difficult and several previous attempts have been unsuccessful \citep[e.g.][]{2002MNRAS.330..187C, 2013AN....334..188R}.

		The {{{problem}}} can be outlined as follows: planets in orbit around a star will reflect a portion of the light from its host star. From geometrical arguments (planet size and semi-major axis of the orbit), \cite{1999ApJ...522L.145C} and \cite{1999Natur.402..751C} predicted that the flux ratio between a planet and its host star will be very low, typically inferior to $10^{-4}$ even for the largest planets known at the smallest orbits. Furthermore, \cite{1999ApJ...513..879M} showed that planetary albedos are extremely dependent on the the atmosphere composition and cloud distribution. {{The giant planets in the Solar System have high albedos (over 0.5) because of condensed molecules in their atmospheres, but planets with temperatures above 400K should have lower albedos, with values between 0.05-0.4 }} \citep[e.g.][]{2011ApJ...729...54C,2008ApJ...689.1345R}. Combining the above factors with the planet's orbital phase as it moves around its host star, the planet becomes virtually undetectable as its signal will have the same order of magnitude as the noise of spectra with signal-to-noise ratio (hereafter S/N) up to to $10^{4}-10^{5}$. 

		In this paper we explore the possibility of using high resolution spectra, combined with the use of the Cross-Correlation technique, to detect the stellar light reflected on the short period giant planets. To do so, we use HARPS spectra to simulate prototypical observation cases and test our methodology. In Sect. \ref{sec:Principle} we describe the principle behind our method and make a brief introduction to the Cross Correlation Function. Section \ref{sec:Application} explains the method and its application to selected cases and in Sect. \ref{sec:Results} we present our results. We discuss them in Sect. \ref{sec:Discussion} and conclude in Sect. \ref{sec:Conclusions} with the lessons learned from our approach.

	\section{The principle}
		\label{sec:Principle}
		The Cross-Correlation Function (henceforth CCF) of a spectrum with a binary mask has been used extensively for the determination of precise radial velocities (e.g. with the HARPS spectrograph)\footnote{Interestingly, \protect\citet{2007A&A...469..355R} have shown that the CCF can be used together with direct imaging techniques to improve our capability to detect small-mass planets.}. A detailed mathematical description of the Cross Correlation Function implementation can be found on \cite{1996A&AS..119..373B} and will not be replicated here. This technique corresponds to stacking a large amount of spectral lines from a spectrum and the resulting CCF single spectral line (named CCF as well in a common abuse of notation) can be seen as an average spectral line of the original spectrum. Assuming that all lines have equal weight or that the weight is taken into account optimally \citep[see][]{2002A&A...388..632P}, the resulting S/N for a given CCF is given simply by
				\begin{equation}
					S/N_{CCF} = \sqrt{n}\, S/N_{spectrum}
					\label{eq:S/NCCF}
				\end{equation}
		where $n$ is the number of spectral lines in the mask used for the CCF. Typically, the binary mask applied to high resolution spectra of a solar-type star has thousands of absorption lines, and thus the S/N on the CCF is much higher than that of the original spectrum. For illustration, using for correlation a stellar mask with 3600 spectral lines, the S/N of the CCF will be 60 times larger than on the spectra. These {{{results show}}} the big advantage of using the CCF over the observed spectra and makes the Cross-Correlation Function a particularly powerful tool for the detection of exoplanets. Until now it has been extensively used to calculate the radial velocity of an object with very high precision; we now propose to use the very high S/N CCF to detect the minute planetary signal.

	\section{Application to HARPS spectra}
		\label{sec:Application}
		To test our capability to detect reflected light from a planet, we performed a set of simulations based on real spectra (Sect. \ref{sec:Data}) with the goal of reproducing very high S/N observations of a star + planet system. Each simulation can be divided in two steps:
		\begin{inparaenum}[i)]
			\item create very high S/N CCFs to mimic observations of the planet around the host star and
			\item apply the procedure to detect the planet's reflected light.
		\end{inparaenum}

		For the creation of high S/N CCFs, we co-added multiple HARPS observations in order to obtain very high S/N spectra and ran them through the HARPS DRS pipeline to generate the corresponding CCFs. The planet + star observations were created by combining these high S/N CCFs to simulate different observations in distinct points of the orbit (Sect. \ref{sec:MethodDataSimulation}).

		For the detection itself, we extract the planet signal from the CCFs by
		\begin{inparaenum}[1)]
			\item dividing them by a template to remove the star's signal and
			\item summing the resulting CCFs, after correction for the planet radial velocity, to stack the planetary signal (Sect. \ref{sec:Detection}).
		\end{inparaenum}

		\subsection{{The data}}
			\label{sec:Data}
			In this paper we use publicly available ESO archival spectra of the bright K-dwarf $\alpha\,Cen\,B$, obtained as part of a planet search program (ESO program IDs 60.A-9036(A), 084.C-0229(A) and 085.C-0318(A)). This target was chosen due to availability of high S/N data. In particular, the spectra that were used are part of a long series of measurements performed on two different nights. This type of observing strategy is similar to the one that would be used in a search for the reflected light spectrum. In total, we compiled 91 HARPS spectra , with {{wavelengths}} ranging from 3779.56 $\AA$ to 6912.9 $\AA$, obtained on the nights 2010-03-25 and 2010-05-25. The spectra were reduced using the HARPS pipeline, and have S/N that varies from 241 to 506 on the 50th order. Although high S/N data for this star are also available from other instruments, we decided to adopt HARPS data due to the high stability of this instrument. A detailed description of the spectrograph characteristics can be found on \cite{2003Msngr.114...20M}.

		\subsection{Creating the spectra}
			\label{sec:MethodDataSimulation}
			
			{{To generate data with the required high S/N, we selected a random set of spectra from our sample and co-added them until the desired high S/N was attained.  Each of the selected spectra had been corrected from the Barycentric Earth Radial Velocity variation which can be quite significant (from around -30km/s to 30km/s in extreme cases). Since  we intended to simulate the variability of the star, different sets of spectra were chosen for each observation, ensuring that for each observation the star would be represented by a different spectrum with a different S/N.}} The planet spectrum was created similarly by combining all spectra from the night 2010-03-25 (38 spectra) in order to increase the S/N. As the planet/star flux ratio is extremely low (as discussed in the Introduction, inferior to $10^{-4}$), the planet's spectrum noise will be negligible compared to the stellar noise. Consequently, we decided to use the same high S/N spectrum to represent the planet on all simulated observations instead of {one spectrum} per observation as done for the star.

			Each of the star + planet observations is constructed by co-adding their signals, shifted by their corresponding radial velocities. Due to the Cross Correlation Function linearity, it is mathematically equivalent to simulate the star + planet's observations by summing both their CCFs or co-adding their original spectra and then compute the CCF of the sum. {{The CCF sampling step is only limited by the step used in its construction (which we control), whereas the spectral numerical resolution (often referred to as sampling) of the original spectrum is limited by the pixel density of the spectrograph (fixed). This is particularly important when we apply a specific radial velocity shift to the signal. In those cases, an interpolation between consecutive pixels is required and the higher the sampling is, the more precise is the operation. By carefully selecting the step in the creation of the CCF, we can minimize the errors when applying a given radial velocity shift to an observation (please note that increasing the CCF sampling step also increases its computing time)}}. Therefore, we decided to create our observations of the star+planet by summing their CCFs instead of their spectra.

			The planet's orbit was assumed as circular\footnote{Although an elliptical orbit could have been used, the ubiquity of short period {{circularised}} planets led us to assume $e=0$ for the sake of simplicity.} and {{modelled}} by Equations \ref{eq:OrbitPhase} to \ref{eq:RVTotal} \citep[see][]{2011MNRAS.415..673L,1999ApJ...522L.145C}.

\begin{figure}
				\includegraphics[width=\columnwidth]{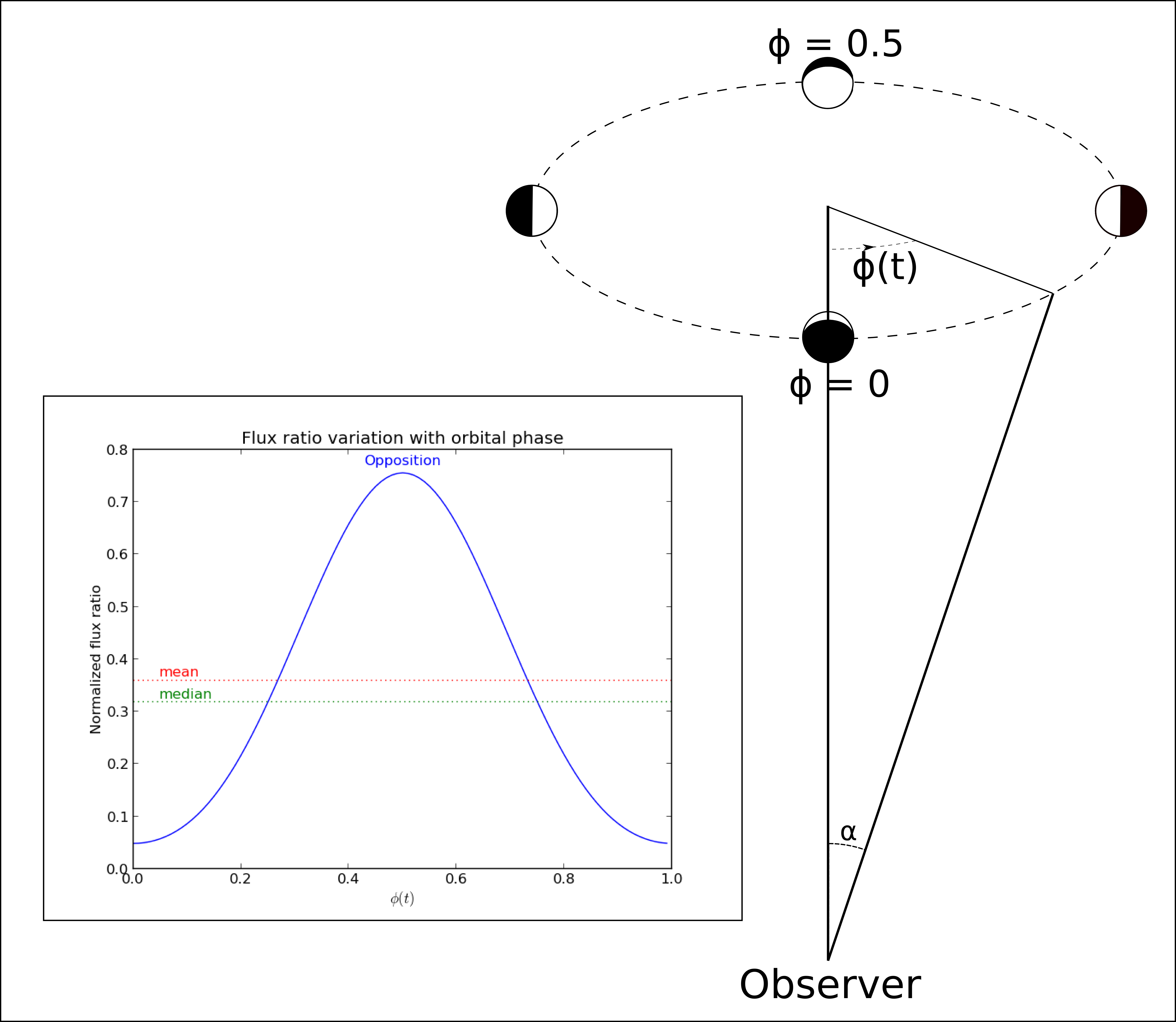}
				\caption{The figure represents the variation of the phase angle {{across}} a full orbit. The inner plot shows the variation of a planet's reflected light as a function of the orbital phase $\phi$ for an inclination of $\pi/2$. The flux ratio has been {{normalised}} by the maximum possible flux for the orbital configuration (orbit edge-on and opposition). The two horizontal lines indicate a value of the mean and median fluxes across one orbit.}
				\label{fig:PhaseAngle}
			\end{figure}

			Assuming that $t_{0}$ is the point of maximum proximity of the planet (the transit epoch) for a given time $t$, the orbital phase $\phi$ of the planet is given by
			\begin{equation}
				\label{eq:OrbitPhase}
				\phi = \frac{t - t_{0}}{P_{orb}}
			\end{equation}
			where $P_{orb}$ is the orbital period of the planet. Given the relative masses of the star and the planet {{$q = M_{planet}/M_{star}$}}, the planet's orbit semi amplitude $K_{planet}$ is given by
			\begin{equation}
				\label{eq:SemiAmplitude}
				K_{planet} = \frac{2 \pi a}{P_{orb}} \frac{sin (I)}{1+q}
			\end{equation}
			where $I$ is the inclination of the orbit relative to us and $a$ the semi-major axis of the planet's orbit. Knowing $\phi$ and $K_{planet}$, the planet radial velocity relatively to the system's barycenter ($RV_{planet, barycenter}(\phi )$) is simply given by
			\begin{equation}
				\label{eq:RVPlanet}
				RV_{planet, barycenter}(\phi ) = K_{planet} sin (2 \pi \phi )
			\end{equation}
			{The radial velocity motion induced on the star by the planet} will be given by
			\begin{equation}
				\label{eq:radial velocityStar}
				RV_{star, barycenter}(\phi ) = - K_{star} sin (2 \pi \phi )
			\end{equation}
			where
			\begin{equation}
				\label{eq:KStar}
				K_{star} = q K_{planet}
			\end{equation}
			Thus, moving to a referential {{centred}} on the star ($RV_{star} = 0$), the planet will have a radial velocity $RV_{planet}(\phi )$ given by
			\begin{equation}
				\label{eq:RVTotal}
				RV_{planet}(\phi ) = RV_{planet, barycenter}(\phi ) + RV_{star, barycenter}(\phi )
			\end{equation}
			For each point of the orbit (at time $t$), the planet's CCF was added to the stellar, shifted by its corresponding radial velocity ($RV_{planet}$) and multiplied by a factor corresponding to its reflected flux relatively to the star. This multiplicative factor has 3 components:
			\begin{description}
				\item[System geometrical configuration]- This component represents {the effective cross-section of the planet relative to the star's incident light ($F_{Geom}$)} and is given by
				\begin{equation}
					\label{eq:Geometrical}
					F_{Geom} = \left[\frac {R_{planet}}{a} \right ]^{2}
				\end{equation}

				\item[Albedo]- The planet's albedo is a measure of the fraction of light its atmosphere/surface will reflect. There are multiple definitions of Albedo, but for simplicity and with no loss of generality, we selected the Geometric Albedo for our simulations \citep[e.g.][]{1999ApJ...513..879M} . The Geometric Albedo $p$ is defined as the reflectivity of a planet measured at opposition, as given by
				\begin{equation}
					\label{eq:BondAlbedo}
					p = \frac{F_{reflected}}{F_{incident}}
				\end{equation}
				which is {{wavelength}} independent and has values between 0 (no reflection) and 1 (full reflection). $F_{reflected}$ and ${F_{incident}}$ are respectively the reflected and incident fluxes.

				\item[Phase function]- Only the planet's hemisphere facing its host star will reflect light. Due to the planet orbital motion and our fixed vantage point, we are not always able to see the whole illuminated hemisphere, but a fraction of it, limiting the reflected flux we receive from the planet. Thus the planet's reflected flux will be dependent on the phase angle $\alpha$ and is {{modelled}} by the phase function $g(\alpha)$ (Equation \ref{eq:PhaseFunction}). The phase function is a Lambert phase function, i.e. assumed for a Lambert Law sphere \citep{2011MNRAS.415..673L}
				\begin{equation}
					\label{eq:PhaseFunction}
					g(\alpha ) = \frac{[sin(\alpha ) + (\pi - \alpha ) \, cos(\alpha)]}{\pi}
				\end{equation}

				where $\alpha$ is the phase angle. Figure \ref{fig:PhaseAngle} shows the relation between the phase angle $\alpha$ and the orbital phase $\phi$, given mathematically by
				\begin{equation}
					\label{eq:PhaseAngle}
					cos(\alpha ) = sin(I) \, cos(2 \pi \phi )
				\end{equation}
			\end{description}

			Combining the different terms, we can easily get the ratio of fluxes from the planet ($F_{planet}$) relatively to the star ($F_{star}$) through
			\begin{equation}
				\label{eq:FluxesRatio}
				\frac{F_{planet}(\alpha)}{F_{star}} = p \, g(\alpha) \left[\frac {R_{planet}}{a} \right ]^{2}
			\end{equation}
			{{A detailed description of equation \ref{eq:FluxesRatio} and its aforementioned components can be found in \cite{2002MNRAS.330..187C}.}} 
			
\begin{figure}
				\includegraphics[width=\columnwidth,page=2]{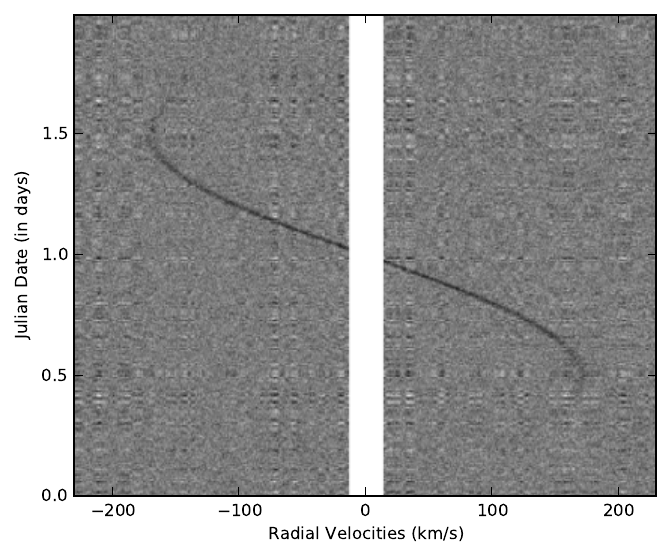}
				\caption{Simulated orbit for a Jupiter mass planet around a one solar mass star ($P_{orb} = 2$ days), {centred} around the stellar radial velocity. The x-axis shows the variation in flux for each system CCF (darker means higher flux) across a radial velocity range and the y-axis  corresponds to {{Julian Date of the simulated observation (the time of the transit has been set to $JD = 0$ for simplification)}}. The planet's orbital movement around the star (the star portion of the CCFs have been excluded from the analysis according to the text and appears as a vertical white strip) can be easily seen on the plot when on opposition (dark path in the middle), but we are unable detect it when it is facing us. \textit{NOTE: this image does not show the raw generated CCFs, but after {normalisation} by the star template.}}
				\label{fig:FullOrbit}
			\end{figure}

			For illustration, if we consider a Jupiter size planet with a 2 day orbit and a wavelength independent albedo of 0.3, the maximum flux ratio is given by
			\begin{equation}
				\label{eq:ExampleJupiter2Days}
				\frac{F_{planet}}{F_{star}} = 0.3 \times 1 \times \left[\frac {69\,911\,km}{4\,637\,534\,km} \right ]^{2} \approx 6.8 \times 10^{-5}
			\end{equation}
			For the same example, the flux ratio variation across a full orbit can be seen on Fig. \ref{fig:FullOrbit}.
			
			For comparison, in the case of the Earth, assuming a fixed albedo of 0.29 \citep[][chap.~3]{2010plsc.book.....D} this ratio is given by
			\begin{equation}
				\label{eq:ExampleEarth}
				\frac{F_{planet}}{F_{star}} = 0.29 \times 1 \times \left[\frac {6\,371\,km}{149\,597\,871\,km} \right ]^{2} \approx 5.3 \times 10^{-10}
			\end{equation}

		\subsection{Detecting the signal}
			\label{sec:Detection}
			The first step in our process was to correct each CCF of the star radial velocity, setting the origin of the referential at the star radial velocity on each CCF. Once all of the star CCFs are aligned at 0km/s, we can proceed with the removal of the stellar signature, a procedure highly dependent on the construction of an accurate template for the star. This template needs to represent the most exactly the star CCF, as it will be used to remove the stellar signal from our observations.

			Instead of using a model, we decided to create a template CCF from our observations; the use of a fully artificial template would make the detection capability extremely dependent on the accuracy of the model and thus much harder to evaluate. To create the template we co-added all the simulated CCFs. The CCFs include both the star and planet reflected spectra along the different orbital phases, which is a situation similar to real observations: these will have some contribution from the planet's reflected light amidst the stellar noise. Constructing the template in this manner, not only increases the S/N of the template, but also dilutes the planet contribution along a radial velocity range, which allows for the construction of an accurate stellar template. Before constructing the template, each CCF was {normalised} by dividing {it} by a second degree polynomial fit to remove small variations in its continuum.  While at first sight there might be no reason to consider that the continuum slope might vary (even if slightly) there are several effects that can contribute to this. The first is the fact that atmospheric absorption is a wavelength-dependent phenomenon, and even if corrected a small residual might remain. The same is true for the instrument transmission as a function of wavelength. These chromatic effects have an important consequence: as our planet travels in its orbit, our Earth's radial velocity will be imprinted in the recorded spectra, shifting them in wavelength. The interplay between the shift in wavelength and the variable transmission as a function of wavelength will lead to a continuum slope variation which has to be corrected before we co-add spectra obtained at different times, which will be done later.

			As discussed before, the stellar signal is several orders of magnitude larger than the planetary one, and as a consequence must be removed first. Two different methods can be used for its removal: i) subtract the star template CCF from the CCFs and ii) {normalise} each CCF by the star template CCF. The subtraction of the template requires that both the template and the CCF are {normalised} to the same flux, or the difference in flux will introduce a residual that adds to the noise, masking the planet's signal. On the other hand, {normalisation} by the template permits to effectively remove the shape of the star signal, regardless of the relative fluxes of both the template and CCF. These relative fluxes will only translate into a multiplicative factor applied to the noise and planet CCF and do not affect their relative intensities. For this reason we opted to {normalise} the CCFs by the template to remove the stellar contribution.

			Note that the S/N of the template is much higher than the S/N from the observations CCFs - for instance, by combining 100 CCFs we will boost the template S/N tenfold - and thus will not reduce the SNR of our observations significantly as we divide by it.

			The removal of stellar signal is bound to be imperfect due to slight mismatches between observed and template CCFs, our {normalisation} is not an exception and will always leave residuals. These residuals are, in first approximation, proportional to the signal, and will be particularly intrusive in a radial velocity range close to the radial velocity of the star. For this reason, detections around opposition  are likely to become unreliable. This is supported by the results of \cite{2013ApJ...767...27B}, where observations around opposition lead to a non-detection in one of the three nights, while the planet was confidently detected on the other two.

			To circumvent this problem, and limit contamination by the star signal, observations close to opposition, where planetary and stellar signals are spectroscopically blended or close, are discarded. Although this will eliminate the most {favourable} cases in what concerns flux, it allows for a clear separation between the star and planet's CCF in radial velocity. By considering that each CCF has an approximate width of $4\,FWHM_{Star}$ \footnote{Since no rotation profile has been introduced on the planet signal, we can assume the width of both planetary and stellar CCFs to be the same.}, we can then safely assume that both CCFs are separated by $4\,FWHM_{Star}$ so they do not contaminate each other. Adding to that $3\,FWHM_{Star}$ on each side of the planet CCF to have enough of a {continuum} for comparison with the planet signal, we discarded the CCFs where the planet and the star have radial velocities inferior to $7\,FWHM_{star}$, which is very conservative.

			\begin{figure}
				\includegraphics[width=\columnwidth,page=4]{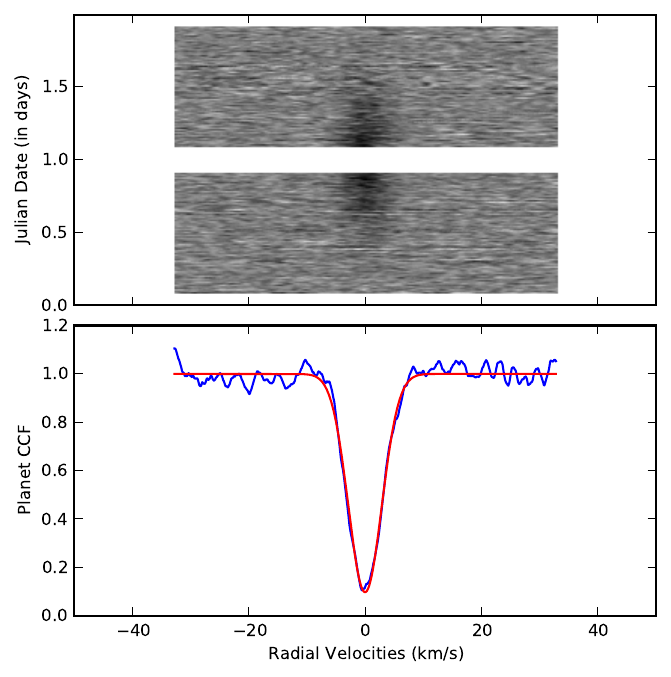}
				\caption{\textit{Top Panel}: Same case as Fig. \ref{fig:FullOrbit}, after correction of the planet's radial velocity and removal of the discarded sections and spectra. The white areas correspond to the discarded spectra. \textit{Bottom Panel}: Integration of the planet's signal over CCFs defined in Fig. \ref{fig:FullOrbit}. This stacking is done by collapsing the CCFs sections in the top panel along the \textit{y} axis.}
				\label{fig:SummedCCF}
			\end{figure}

			After {normalisation}, we correct each CCF for the planet's radial velocity. Given the time of each observation and the orbital parameters of the planet's orbit, Equation \ref{eq:RVPlanet} will deliver the radial velocity of each CCF.

			Finally, we co-add the resulting {normalised} CCFs, which increases the detectability of the planet signal by a factor proportional to the square root of the number of CCFs. For the sake of simplicity, we name the co-added spectra $CCF_{stack}$.

			To determine the significance of the detection, a {Gaussian} curve, defined by equation \ref{eq:GaussFit}, is fitted to $CCF_{stack}$:
			\begin{equation}
				\label{eq:GaussFit}
				y = B + A \cdot e^{-2 \ln 2 \left(\frac{x - mean}{FWHM}\right)^2}
			\end{equation}

			where $A$ (Amplitude), $FWHM$, $B$ (ordinate for the continuum) and $mean$ are the free parameters of the Gaussian fit. The significance of the detection is then given by the ratio of the amplitude of the {Gaussian} fit by the noise. The noise is given by the residuals around the {continuum} fit. Fig. \ref{fig:PhaseAngle} shows the application of this technique to the detection of a Jupiter size planet with a 2 day orbital period.

			Several constrains were set to eliminate spurious Gaussian fits with no physical meaning. First of all, the amplitude is constrained by assuming that its value is strictly negative (a positive value for the amplitude would imply a negative albedo).

			The FWHM is the parameter that requires more careful considerations. Since we consider that the light from the planet is reflected light from the star, its CCF should replicate the one from the star, corrected to account for physical mechanisms that might widen the spectral lines. In our simulations, no deformation (e.g., planetary rotation) was applied to the star's CCF to create the planet's CCF, hence both CCFs have the same $FWHM$. However, since our CCFs are affected by noise, the planet's CCF will not have the same $FWHM$ as the star's. A lower limit of $0.75FWHM_{Star}$ to the planet's $FWHM$ is imposed assuming that no physical process can reduce the CCF width, but taking into account a 25\% margin for flexibility of the {minimising} algorithm. A higher limit is set dynamically in our simulations assuming a rotational broadening of the CCF due to the planet being in tidal lock with its host star \cite[e.g.][]{1997ApJ...484..866L} as a result of its close proximity to the star, and consequently low orbital period. To this broadening a 25 \% margin is added for flexibility of the {minimising} algorithm\footnote{The broadening of the CCF can be estimated from $v_{rot}\,sin\,i = A\sqrt{FWHM_{rot}^{2} - FWHM_{0}^{2}}/2.35$  where $v_{rot}$ is the rotational velocity, $i$ the rotation axis inclination, $A$ the coupling constant, $FWHM_{0}$ and $FWHM_{rot}$ are the natural and after rotation widths of the CCF \citep[see][Chapter 18]{2008oasp.book.....G}. As an example, assuming a value of 1.9 for the coupling constant \citep{2002A&A...392..215S}, that $FWHM_{0} = FWHM_{Star} \approx 6.6$ (average value for the CCFs FWHM) and an edge-on orbit, in the case of a 10 hour orbit Jupiter sized planet we will have $FWHM_{rot} \approx 1.08 FWHM_{0}$.}.

			Both the $mean$ and the $B$ parameters were left unconstrained.

	\section{Results}
		\label{sec:Results}
		\subsection{Cases of study}
		\label{sec:Cases}
			Applying the methodology to the data described in Sect. \ref{sec:Application}, we simulate the detection of different planets. Our cases of study are intended to mimic data as obtained with two instruments: the future ESPRESSO spectrograph for the VLT \citep[Echelle SPectrograph for Rocky Exoplanets and Stable Spectroscopic Observations,][]{2010SPIE.7735E..14P}, and a high resolution spectrograph (such as the proposed HIRES) for the E-ELT. {{We used UVES \cite[Ultraviolet and Visual Echelle Spectrograph,][]{2000SPIE.4008..534D} Exposure Time Calculator\footnote{\protect\url{http://www.eso.org/observing/etc/bin/gen/form?INS.NAME=UVES+INS.MODE=spectro}} to compute the S/N for a given exposure time for the VLT simulations. For the E-ELT simulations, we used the same tool and extrapolated the results by multiplying by a factor of 4.8, the approximate diameter ratio of the E-ELT mirror and one of VLT’s primary mirrors (respectively around 39 and 8.2 meters).}} The resulting data was then used to simulate the detection of both theoretical and real planetary systems selected from the encyclopedia catalog Exoplanet.eu \citep{2011A&A...532A..79S}.

			Our cases of study consist in planets representative of two different planet populations found around other starts: Jupiter and Neptune   planets in close orbits (see table \ref{tab:SimulatedPlanets}). Furthermore, we selected some real examples from the exoplanets.eu database. As we require a high S/N, we selected 51 Peg b and 55 Cnc e, as both orbit bright stars. They are also both good representatives of hot Jupiter and super-Earth planet types respectively. Except for 55 Cnc e where the radius has been determined\citep[see][]{2012A&A...539A..28G}, the radii for the simulated planetary systems has been extrapolated from planets with similar masses described in \url{http://exoplanets.eu} (See table \ref{tab:SimulatedPlanets} for details). For each case of study, we simulated detections assuming observations distributed evenly across two 5 hour periods on each side of phase $\phi = 0.5$ (opposition), encompassing the most {favourable} orbital phases (higher planet/star flux ratio). {{For all cases we considered a wavelength independent albedo of 0.3, which can be considered optimistic as the short period planets we simulated should be cloud free and therefore have albedos of 0.05 - 0.4 as referred before \citep[e.g.][]{1999ApJ...513..879M}.}}

			\begin{table*}
				\begin{minipage}{\textwidth}
				\caption{Simulated planets data.}
				\resizebox{\textwidth}{!} {
					\begin{tabular}{l c c c c c c c c c c c}
						\hline\hline\\
						Planet		&	Period	&	Planet Radius	&	Planet Mass		& 	Maximum	&	\multicolumn{2}{c}{Exposure}&	\multicolumn{2}{c}{Number of}& \multicolumn{2}{c}{Expected Detection}	\\
						Name		&	(days)&	(Earth radii)& 	(Earth masses)& 	Flux Ratio&	\multicolumn{2}{c}{Time (seconds)}&	\multicolumn{2}{c}{of {{Exposures}}}&	\multicolumn{2}{c}{Significance ($\sigma$)}\\
										&			&			& 			&							&	ESPRESSO	&	HIRES	&	ESPRESSO	&	HIRES	&	ESPRESSO		&		HIRES		\\
						\hline{}\\[.1ex]
						\multicolumn{6}{l}{{Prototypical Planetary Systems:}}	&	\\[1.1ex]
						Jupiter		&	3	&	13$^{\textit{a}}$	&	319	&	$5.3 \times 10^{-5}$	&	900	&	--	& 	40	&	--	&	{34}	&		\\[1.1ex]
						Neptune		&	2	&	5.5$^{\textit{b}}$	&	17	&	$1.6 \times 10^{-5}$	&	900	&	600	& 	40	&	60	&	 {10}	&	 {19}	\\[1.1ex]
						\multicolumn{6}{l}{{Real Planetary Systems:}}	&	\\[1.1ex]
						51 Peg b	&	3.09	&	10$^{\textit{c}}$		&	149.29	&	$1.8 \times 10^{-5}$	&	600	&	60	&	60	&	600	&	{15}	&	{69}		\\[1.1ex]
						55 Cnc e	&	0.74	&	2.04	&	8.39	&	$9.1 \times 10^{-6}$	&	900	&	90	&	40	&	400	&	4	&	{20}	\\[1.1ex]
						\hline\hline

					\end{tabular}}
				\medskip{}\\
				{For the real planetary systems, the data was obtained from \protect\url{http://exoplanets.eu} and rounded to the second decimal place. The Maximum Flux Ratio corresponds to the planet/star flux ratio at opposition, i.e. when reflection is {maximised}, assuming an albedo of 0.3. The Exposure Time is the expected integration time required to obtain a S/N of 1600 with the ESPRESSO and 2400 with the HIRES. For the Expected Detection Significance, we consider a total exposure time of 10h and compute it taking into account the phase variation along the orbit using Equation \ref{eq:ExpectedDetection}. The radii for all the simulated case \citep[except 55 Cnc e where the radius has been determined, see][]{2012A&A...539A..28G} has been extrapolated from planets described in \protect\url{http://exoplanets.eu} that have similar masses:
				$\itl{a)}${ Radius from WASP-48 b}; $\itl{b)}${ Radius from Kepler-18 c}; $\itl{c)}${ Radius from WASP-60 b}}
				
				\end{minipage}
				\label{tab:SimulatedPlanets}
			\end{table*}

			To create our CCFs, we used a step of 0.1 km/s, the mask for a K1 spectral type star (with {4839} spectral lines), a width of 320 km/s and a target radial velocity of -22 km/s. For all other settings we used the default values of the pipeline.

			\subsection{Results of the simulations}

			For the different scenarios, we have a rule of thumb that the expected detection significance for an observational data set is given by
			\begin{equation}
				\label{eq:ExpectedDetection}
				Detection_{expected} = \overline{S/N}_{spectra} * \sqrt{N_{CCF} * \sum_{i}\left(\frac{F_{planet}}{F_{star}}\right)_{i}^{2}}
			\end{equation}
			in which the lower script $"i"$ corresponds to the different observations considered along the orbit. $N_{CCF}$ represents the number of lines in the mask used to create the CCF, $\overline{S/N}_{spectra}$ is the average S/N of the original spectra and $F_{planet}/F_{star}$ is the flux ratio between the planet and the star as obtained from Equation \ref{eq:FluxesRatio}. In our cases of study, we considered $N_{CCF}=4120$, the number of lines defined in the mask of a K1 star on the HARPS DRS pipeline.

			We consider a detection to be successful (or significant) when we are able to perform a Gaussian fit to the planet's CCF, whose amplitude modulus is larger than 3 times the noise, with the latest being the standard deviation of the points in the continuum.

\begin{figure}
				\includegraphics[width=0.45\textwidth,page=5]{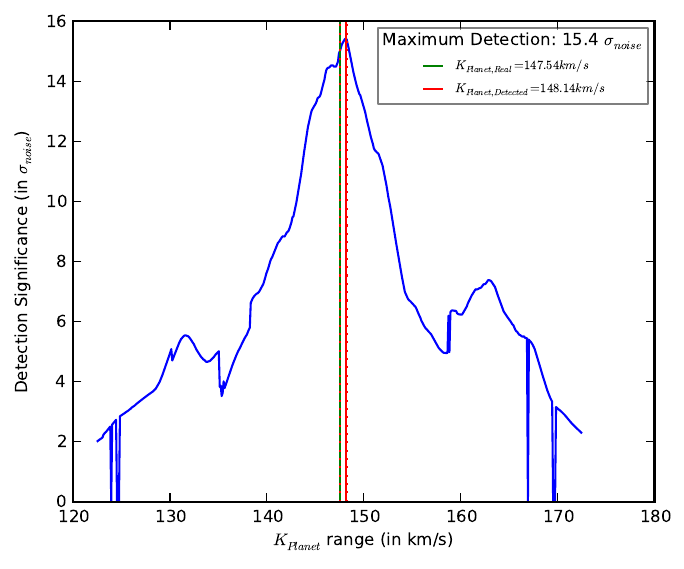}
				\caption{Variation of the Detection Significance across a range of radial velocity values for the semi-amplitude. This case represents a simulation for a Jupiter type planet with a 3 day orbital period as observed with the VLT. It can be seen that the Detection Significance decreases as we move further from the correct orbital semi-amplitude $K_{Real}$. Also, the maximum value of the Detection Significance occurs close to the correct orbital semi-amplitude. This is in agreement with what we expected: for a semi-amplitude close to the real value, the planet signals get summed together, increasing the detection significance. For semi-amplitudes where the {minimising} function was unable to fit a Gaussian curve whose parameters are within our constrains, the value of the detection significance was set to zero, which shows as gaps in the plot.}
				\label{fig:AmplitudeVariation}
			\end{figure}

\begin{figure}
				\includegraphics[width=\columnwidth,page=1]{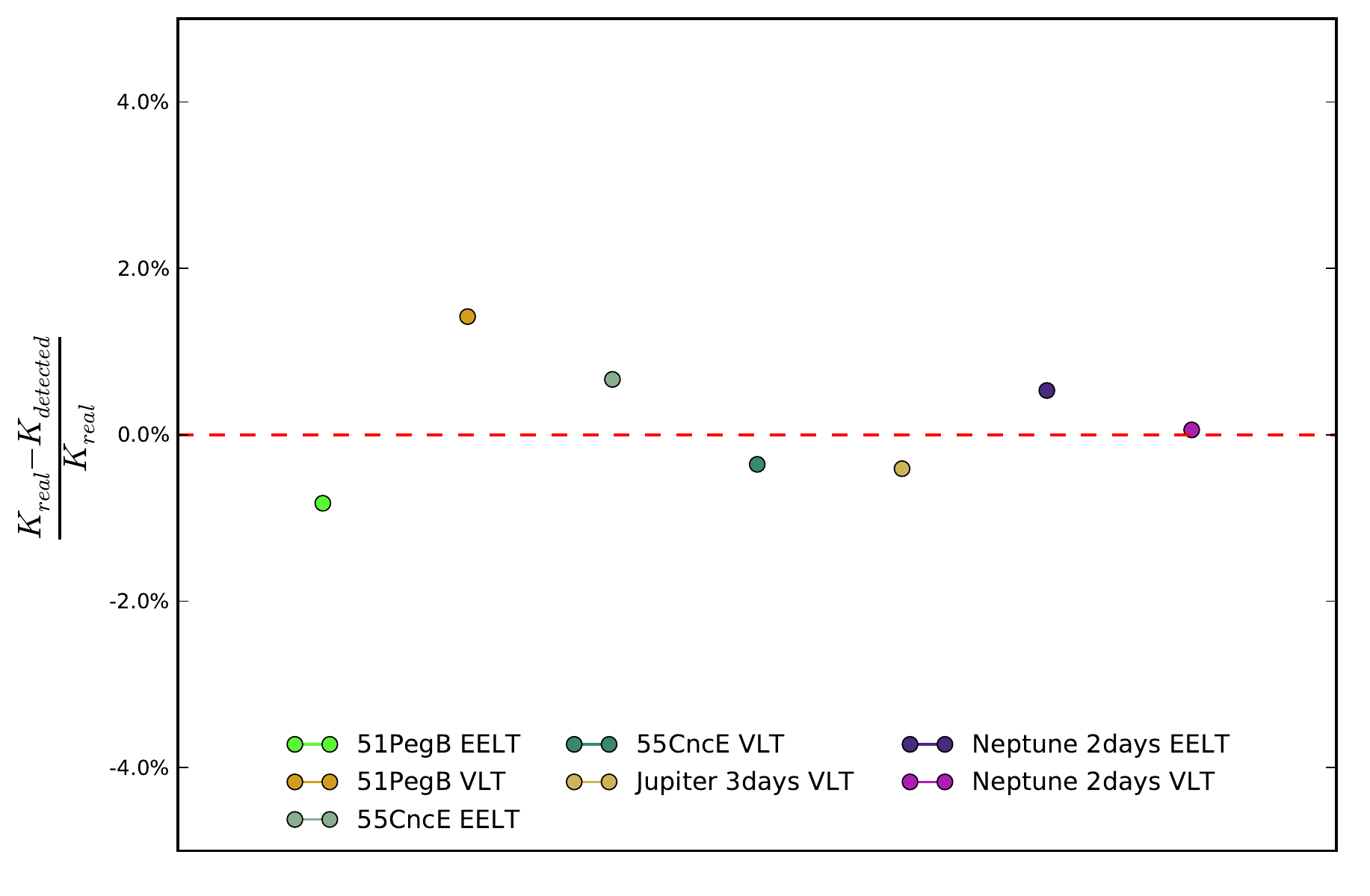}
				\caption{Plot of $K_{real} - K_{detected}$ for each case of study. The horizontal {solid} red line represents the ideal case, where $K_{detected} = K_{real}$. It can be seen that in the cases where we have a significant detection, the difference between the simulated ($K_{real}$) and detected ($K_{detected}$) semi-amplitudes is inferior to 2\%.}
				\label{fig:KErrorsGraph}
			\end{figure}

			To confirm that we were detecting the stacking of the planet's CCFs and not random noise artefacts, we tested the detection of the planet's signal along a range of orbital semi-amplitudes {centred} on the real planetary semi-amplitude $K_{real}$ and with a width of 50 km/s. By construction, the maximum detection significance should correspond to $K_{real}$. Defining $K_{detected}$ as the detected significance minimum, we should have $K_{Detected} = K_{Real}$, which corresponds to the situation where the planet's signal is perfectly aligned on all CCFs and we are fitting the signal of the correct orbit to our data. As we move further away from $K_{real}$, we can notice an effect of smearing of the multiple CCFs with the reduction of the significance, as instead of adding the centre of the CCFs, we will be adding their wings (see Fig. \ref{fig:AmplitudeVariation}). Therefore we will also be able to fit a Gaussian to the sum of the CCFs, but with a larger FWHM and lower amplitude, and consequently a lower detection significance. This effect gets more noticeable as we move further from $K_{real}$, until the CCFs stop overlapping altogether. Since the CCFs have an average $FWHM$ of around 6.6 km/s, this overlapping stops to be significant at around $2 \times FWHM_{Star}\approx 13.2 \,km/s$. At this point, we should have a very low detection significance. However, the presence of systematic noise (e.g., the vertical stripes in figure \ref{fig:FullOrbit}), will impart a variation to the noise (measured by the standard deviation) much larger than random noise only. Since the detection significance as we defined it depends on the standard deviation of the fit to the continuum, this variation will affect the detection significance considerably. Therefore, the maximum detection significance might not happen for $K_{Real}$. The {{{difference}}} $K_{real} - K_{detected}$ for each of our cases of study is shown on Fig. \ref{fig:KErrorsGraph}.

\begin{figure*}
				\centering
				\subfigure{\includegraphics[width=\columnwidth,page=1]{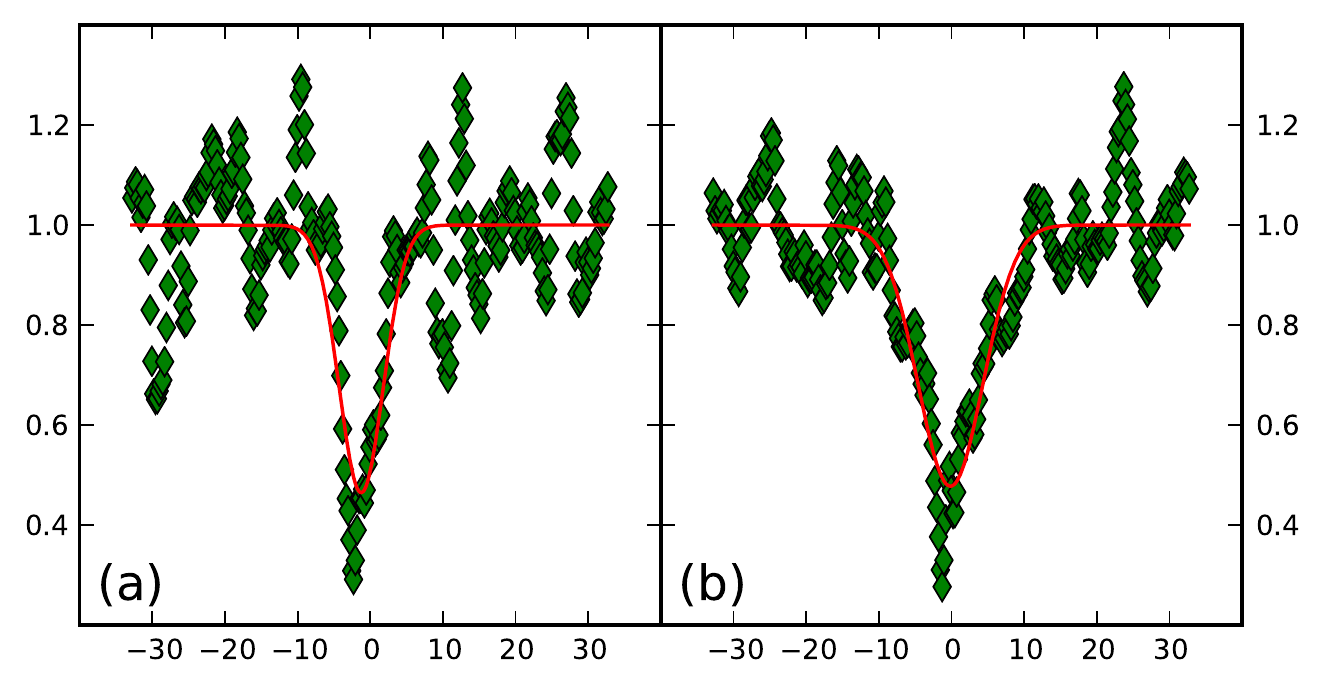}}
				\subfigure{\includegraphics[width=\columnwidth,page=2]{./figures/ResultsPlot.pdf}}
				\subfigure{\includegraphics[width=\columnwidth,page=3]{./figures/ResultsPlot.pdf}}
				\subfigure{\includegraphics[width=\columnwidth,page=4]{./figures/ResultsPlot.pdf}}

				\caption{Results from simulations. In each panel, we can see the extracted planet CCF for each case (green diamonds), as well as the best Gaussian fit (red line). In each case we were able to extract the planetary signal successfully and perform a significant detection. The continuum of each CCF was normalised to 1 and the CCFs are not to scale with each other. 
					\textit{(a)} Neptune (VLT), 			Period = 2.0 days,		S/N = 4.4;
					\textit{(b)} Neptune (E-ELT), 		Period = 2.0 days,		S/N = 5.9;
					\textit{(c)} 51 Peg b (VLT), 		Period = 4.2 days,		S/N = 5.2;
					\textit{(d)} 51 Peg b (E-ELT), 		Period = 4.2 days,		S/N = {24.6};
					\textit{(e)} 55 Cnc e (VLT), 		Period = 0.7 days,		S/N = 2.7;
					\textit{(f)} 55 Cnc e (E-ELT), 		Period = 0.7 days,		S/N = 4.9;
					\textit{(g)} Jupiter (VLT), 			Period = 3.0 days,		S/N = 15.4}
				\label{fig:Results}
			\end{figure*}

			The results of our simulations for each case of study can be found on Table \ref{tab:ResultsPlanetSelection} and Fig. \ref{fig:Results}. For both the prototypical planets we were able to recover the planetary signals with above a $3\sigma_{noise}$ significance (Jupiter with VLT: $15.4\sigma_{noise}$; Neptune with VLT: $4.4\sigma_{noise}$; Neptune with E-ELT: $5.9\sigma_{noise}$). For the real planetary cases, 51 Peg was recovered with both telescopes (VLT: $5.2\sigma_{noise}$; E-ELT:$20.3\sigma_{noise}$), but 55 Cnc e signal was only recovered with the E-ELT ($4.9\sigma_{noise}$). For 55 Cnc e with the VLT, although the detection is not considered significant (S/N = 2.7 $\sigma_{noise}$), a clear signal can be observed in Fig. \ref{fig:Results}, panel \textit{(e)} where the planet CCF is supposed to be. The results concerning the difference $K_{real} - K_{detected}$ can be found in Table \ref{tab:ResultsPlanetSelection} and Fig. \ref{fig:Results}.

			\begin{table*}
				\begin{minipage}{0.8\textwidth}
					\caption{Results of simulations.}
					\centering
						\begin{tabular}{l c c c c c c c c c}
							\hline\hline\\
							Planet Name	&	$K_{Real}$	&	\multicolumn{2}{c}{$K_{Detected}$}	&	\multicolumn{2}{c}{$|K_{Real} - K_{Detected}|$}	&	\multicolumn{2}{c}{Detection Significance}	\\
								&	&	ESPRESSO	&	HIRES	&	ESPRESSO	&	HIRES	&	ESPRESSO	&	HIRES	\\
							\hline{}\\
							\multicolumn{6}{l}{{Prototypical Planetary Systems:}}	&	\\[1.1ex]
							Jupiter (3 days)	&	147.54	&	148.14	&	--	&	0.60	&	--	&	15.4$\pm$0.3	&	--	\\[1.1ex]
							Neptune (2 days)	&	168.99	&	168.89	&	168.09	&	0.10	&	0.90	&	4.4$\pm$0.3	&	5.9$\pm$0.2	\\[1.1ex]
							\multicolumn{6}{l}{{Real Planetary Systems:}}	&	\\[1.1ex]
							51 Peg b	&	133.77	&	131.87	&	{134.87}	&	1.90	&	{1.10}	&	5.2$\pm$0.3	&	{24.6$\pm$0.4}	\\[1.1ex]
							55 Cnc e	&	225.30	&	226.60	&	223.80	&	0.80	&	1.50	&	2.7$\pm$0.2	&	4.9$\pm$0.1	\\[1.1ex]

							\hline\hline
						\end{tabular}\\
					\medskip{}
					The values in columns $K_{Real}$, $K_{Detected}$ and $|K_{Real} - K_{Detected}|$ are in $km/s$. The Detection Significance in units of $\sigma_{noise}$. For comparison, the expected values can be found on table \ref{tab:SimulatedPlanets}.
				\end{minipage}
				\label{tab:ResultsPlanetSelection}
			\end{table*}

	\section{Discussion}
		\label{sec:Discussion}		
		Although our results show detections with lower amplitude than expected (e.g. 51 Peg b with the VLT, for which we get a detection of around $5.2 \sigma_{noise}$ instead of the expected $13 \sigma_{noise}$), these can be explained by systematic features in the CCFs. These features can be see on Fig. \ref{fig:FullOrbit} as vertical stripes and will add systematic noise to the continuum. Since the detection significance is highly dependent on the noise of the continuum, these systematics will lower the detection significance and mask the planet signal. A clear example of this is shown in Fig. \ref{fig:Results}, panel \textit{f}, the case of 55 Cnc e with the E-ELT, where a clear detection can be seen, but the systematics totally deform the continuum around the planet's signal. We must not forget, that although we are simulating spectra as observed with future instruments and telescopes by increasing their S/N, we are still using data from current observing facilities. Therefore, our simulated spectra will still be affected by the same systematics, but transposed to a higher S/N domain. By construction, these systematics will also increase by a factor proportional to the square root of the number of real spectra used to create the simulated ones. This will most definitely affect the significance of our detections, which will be lower than predicted theoretically, and {{can be considered pessimistic{\footnote{{Although the systematics in the noise lead to pessimistic estimates of the detection significance, we can safely assume that the assumption of a generally optimistic albedo should help balancing this effect.}}}}}. 
		
		{As a side test we combined Equations \ref{eq:FluxesRatio} and \ref{eq:ExpectedDetection} to estimate the planet's albedo from the recovered planetary signal amplitude. In the cases where we had a significant detection we got an albedo estimate 2-3 times smaller than simulated. This result is in accordance with what has been discussed in the previous paragraph about the influence of systematics in the noise.}
		
		It is interesting to note that in all cases, and in spite of the very different S/N on $CCF_{stack}$, the semi-amplitude of the planet orbit is recovered with an associated error of the order of a couple of percent (Fig. \ref{fig:KErrorsGraph}).
		
		Furthermore, in the previous section we chose to determine the uncertainty on the stacked spectra $CCF_{stack}$ by analyzing the spectra itself instead of deriving it theoretically; this is a conservative approach. However, measuring the noise by calculating the standard deviation of the residuals can lead to over-optimistic values when one is in the presence of correlated noise. In particular, if the correlation distance between the points is similar to that of the scale of our signals, the detection will be far more difficult than in the presence of white noise. This leads us to believe that, while widespread, using the standard deviation of the residuals to measure the noise can artificially boost the significance. This is also suggested by Fig. 4, in which we can see that structures are detected with a significance larger than 3 sigma for an orbital semi-amplitude which is offset relative to $K_{Real}$ by several FWHM's and should thus display a non-detection.

		Following this lead, we considered a different approach for the calculation of the noise level. We computed, in $2\times FWHM$ interval pixel interval, the peak-to-peak variation in flux in the continuum of $CCF_{stack}$. We applied this procedure to the $CCF_{stack}$ calculated for $K_{Real}$ value in order to ensure we are measuring the peak-to-peak variations which were created by artefacts and systematics and not the diluted planetary signal. The average of these peak-to-peak variations turned out to be roughly two to three times larger than the noise as {{{calculated}}} in the previous section. This calculus can be seen as a pessimistic view or worst-case-limit of the noise analysis: if the correlated noise or artefact signal is of much larger amplitude than the consecutive pixel-to-pixel variations, the peak-to-peak variations will be the main obstacle to the detection of the signal and thus the indicator that should be considered. Yet, a much finer analysis can in principle allow to distinguish a signal from the noise.

		If we consider the noise to be three times larger than previously assumed, the detections of the prototypical Neptune and 55 Cnc e with both the VLT and E-ELT, as well as 51 Peg b with the VLT, are no longer significant. However, we stress that this approach should be seen as an ad-hoc pessimistic correction. It will depend on all factors that can create a spurious signal or systematic, and cannot be {generalised} easily. In fact, and as discussed previously, even {{{though}}} we tried to do our simulations staying as close to observed data as possible, it is likely that we introduced some artefacts in our analysis. Once more, the take-away lesson is that one should expect spectra  and CCFs to contain artefacts which are 2-3 times larger than the noise as obtained by back-of-the-envelope calculations.
		
		{{One of the assumptions we are making is that the planetary spectrum is an exact copy of the stellar one. Although this is a simplification and the planetary spectrum will surely have more spectral features, this assumption is not expected to have a strong impact on the observations as long as the stellar lines used in the mask are present in the planetary spectrum, and not polluted by planetary lines.}}
		
		In order to {optimise} a planet's reflected light detection, a careful selection of the time and date of our observations in order to get the most {favourable} cases (e.g., between maximum elongation and the limit where the planet's and star's CCFs merge) is required. Otherwise, we risk getting observations of the star + planet when the later is not facing us, e.g. close to transit time, and therefore we will not be able to recover its signal.

	\section{Conclusions}
		\label{sec:Conclusions}
		We simulated observations of prototypical planetary systems (hot Jupiter and Neptune types with respectively 3 and 2 days orbital periods) and real planetary systems (51 Peg b and 55 Cnc e) with both VLT (for all planetary systems) and E-ELT (for all planetary systems except the hot Jupiter) to test the possibility of detecting the planet's reflected light with next generation instruments (ESPRESSO@VLT and HIRES@E-ELT). In these simulations, we were able to to recover the planetary signals with a significance above $3\sigma_{noise}$ (Jupiter with VLT: $15.4\sigma_{noise}$; Neptune with VLT: $4.4\sigma_{noise}$; Neptune with E-ELT: $5.9\sigma_{noise}$; 51 Peg b with VLT: $5.2\sigma_{noise}$; 51 Peg b with E-ELT:$24.6\sigma_{noise}$; 55 Cnc e with the E-ELT: $4.9\sigma_{noise}$). This shows it is indeed possible to detect a planets's reflected light by using the Cross Correlation Function as it allows us to operate in a much higher signal-to-noise ratio domain, by a factor proportional to the square root of the number of CCFs. 
		
		The simulation of the observations of 51 {{{Peg}}} b with the VLT is particularly interesting, as it shows a clear detection for a real object {{{with a}} current telescope, even if with marginal significance. It is also interesting to note that we were also able to recover the correct orbital semi-amplitude {{{with}} an associated error of the order of a couple of percent (Fig. \ref{fig:AmplitudeVariation}{{) for}} all scenarios.
		
		In all simulations, we tried to simulate observations as close to reality as possible, namely by avoiding the use to artificial spectra. By doing this we intended to understand how systematics present in real spectra affect the detection of the planetary signal. This is particularly important as these systematics add to the continuum noise and affect the detection significance of the planet's reflected light.
		
		In an alternative and more pessimistic approach to the noise evaluation, where instead of the standard deviation of the continuum flux we consider a peak-to-peak variation in the flux, resulting from these systematics, the noise variation will be 2-3 times larger. Therefore, the detection significance will decrease by the same factor, turning lower significance detections like 51 Peg b with VLT or Neptune with VLT into non-detections.

		The arrival of new tools and technologies, e.g. ESPRESSO@VLT and HIRES@E-ELT, will permit the collection of spectra with the required S/N improvement, which should in turn be accompanied by the reduction of the aforementioned systematic features and errors. In particular, paramount to this objective are the larger dimensions of next-generation telescopes like ESO's E-ELT, allowing for the collection of an increased number of spectra on the same period of time. This should allow for the detection the reflected light of smaller planets at longer period orbits with increased precision, which in turn will permit the study of planetary atmospheres with more detail.

\section*{Acknowledgments}
	We acknowledge the support by the European Research Council/European Community under the FP7 through Starting Grant agreement number 239953, as well as from Funda\c{c}\~ao para a Ci\^encia e a Tecnologia (FCT) in the form of grant reference PTDC/CTE-AST/120251/2010. Pedro Figueira had further support from Funda\c{c}\~ao para a Ci\^encia e a Tecnologia (FCT) in the form of grant reference PTDC/CTE-AST/098528/2008. {{We thank Francesco Pepe for interesting discussions on the subject which were of great help to take this paper further. We acknowledge the referee, Ronald Gilliland, for a careful review and drawing our attention to important shortcomings of our analysis.}}

\bibliographystyle{mn2e}
\bibliography{bibtex_refs_MNRAS}

\appendix

	\section{Signal to Noise transformation}
		\label{sec:S/NEvolution}
		One of the main concerns through this work was the creation of high S/N spectra in a realistic way and how each operation transforms the S/N of these spectra. We review these transformations here.

				A common technique to increase the S/N of spectra is the addition of multiple spectra. The principle is simple: the signal being a constant in all spectra will add to itself, but white noise will average out. Thus, for the sum of multiple spectra, the S/N will be given by
				\begin{equation}
					\label{eq:S/NSum}
					S/N_{sum}^{2} = \sum\limits_{i} S/N_{i}^{2}
				\end{equation}
				where $S/N_{i}$ and $S/N_{sum}$ are respectively the S/N of each individual spectrum and of their sum. In the particular case where all spectra have similar S/Ns, we can approximate Equation \ref{eq:S/NSum} by:
				\begin{equation}
					\label{eq:CCFS/N}
					S/N_{sum} = \sqrt{N}\,\,S/N
				\end{equation}

				where N is the number of spectra we are summing and $S/N$ is the average S/N of each individual spectrum.

				A completely different issue is the division of two spectra. To verify that the division of two Poisson distributions is still a Poisson distribution, we ran a series of Monte Carlo simulations where we divided two Poisson distributions with high S/N ($S/N > 2000$) by each other. For such high S/N, and as expected, Poisson distributions are similar to Normal distribution with mean $\mu = \lambda$ and standard deviation $\sigma = \sqrt{\lambda}$. Our Monte Carlo simulations showed that the resulting distribution from the division would also be gaussian with parameters $\sigma$ and $\mu$. Defining $Z = \frac{X}{Y}$, being $X$ and $Y$ two arbitrary normal random variables whose distributions have parameters $\mu_{X}$, $\sigma_{X}$, $\mu_{Y}$ and $\sigma_{Y}$, the error propagation formula for measured quantities states that:
					\begin{equation}
						\frac{\sigma_{Z}}{\mu_{Z}} =  \sqrt{\left(\frac{\sigma_{X}}{\mu_{X}}\right)^{2} + \left(\frac{\sigma_{Y}}{\mu_{Y}}\right)^{2}}
						\label{eq:ErrorPropagation}
					\end{equation}
				Since that for each of the variables $X$, $Y$ and $Z$ the signal-to-noise ratio will be given by $S/N_{X,Y,Z} = {\mu_{X,Y,Z}}/{\sigma_{X,Y,Z}} = \sqrt{\mu_{X,Y,Z}}$ and applying equation \ref{eq:ErrorPropagation} to the division of a spectrum $k$ by the star template (as performed in Sect. \ref{sec:Detection}), the resulting S/N for spectrum $k$ ($S/N_{k,Normalized}$) is given by:
					\begin{equation}
						\label{eq:SN1plusN}
						\begin{array} {r l}
						S/N_{k,Normalized} \approx	&	\left(\sqrt{S/N_{k}^{-2} + S/N_{Template}^{-2}}\right)^{-1}\\
						\approx	&	S/N_{k}\left(\sqrt{1 + \frac{1}{N}}\right)^{-1}
						\end{array}
					\end{equation}
				where $S/N_{k}$ and $S/N_{Template}$ are respectively the S/N of spectrum $k$ and of the star template.	Furthermore, for a high enough number of points in orbit/spectra ($>100$), we can safely assume that:
					\begin{equation}
						\label{eq:SNFinal}
						S/N_{k,Normalized} \approx S/N_{k}
					\end{equation}

\end{document}